\documentclass[twocolumn,pra, longbibliography]{revtex4-1}

\usepackage{float}
\usepackage{epsfig}
\usepackage{graphicx}
\usepackage{amsmath}
\usepackage{amssymb}
\usepackage{epstopdf}
\usepackage{xcolor}
\usepackage{bm}
\usepackage{dsfont}
\usepackage{hyperref}
\usepackage{braket}
\usepackage{color}

\begin{document}


\title{Implementation of discrete positive operator valued measures on linear optical systems using CS decomposition}
\author{Jaskaran Singh}
\email{jsinghiiser@gmail.com}
\affiliation{Department of Physical Sciences,
	Indian
	Institute of Science Education \&
	Research (IISER) Mohali, Sector 81 SAS Nagar,
	Manauli PO 140306 Punjab India.}
\author{Arvind}
\email{arvind@iisermohali.ac.in}
\affiliation{Department of Physical Sciences,
	Indian
	Institute of Science Education \&
	Research (IISER) Mohali, Sector 81 SAS Nagar,
	Manauli PO 140306 Punjab India.}
\author{Sandeep K. Goyal}
\email{skgoyal@iisermohali.ac.in}
\affiliation{Department of Physical Sciences,
	Indian
	Institute of Science Education \&
	Research (IISER) Mohali, Sector 81 SAS Nagar,
	Manauli PO 140306 Punjab India.}

\begin{abstract}
Positive operator valued measurements (POVMs) 
play an important role in
efficient  quantum communication and computation.
While optical systems are one of the strongest
candidates for long distance quantum communication and 
information processing, efficient
methods to implement POVMs in these systems are scarce. Here
we propose an all-optical  scheme to implement an arbitrary
POVM using linear optical components on $m$-dimensional
Hilbert space of internal degrees of freedom. Linear optical nature
of the proposed scheme makes it efficient and robust. We
show how the scheme can be applied for state tomography and
for preparing arbitrary mixed states.

\end{abstract}

\maketitle

\section{Introduction}

Projective measurements play an important role in
information theoretic applications of quantum theory.
However, they are not the most general type of measurements
or even the optimal ones in most
cases~\cite{nielsen_chuang,
state_discrimination1}. For example,
in quantum optics, the homodyne and
heterodyne measurements~\cite{gerry_knight,scully} routinely
implemented in the lab cannot be modelled as projective
measurements. The most general class of measurements are
mathematically represented as quantum instruments~\cite{holevo, wiseman,heinosaari, busch} with positive operator valued
measures (POVM)~\cite{nielsen_chuang,aperes} 
being
a special case of them. POVMs and their
applications have been theoretically
studied~\cite{povm_review, povm_app1,povm_app2,povm_app3,sahel_povm}, however their
experimental implemention on optical systems still remains a
challenge. In this paper we propose a scheme  to implement
an arbitrary POVM on optical systems.
The
scheme uses the Naimark dilation theorem and
CS-decomposition~\cite{sandeep_csd} in order to realize the
POVMs.

POVMs have proven to be advantageous in
quantum state
discrimination~\cite{state_discrimination1,state_discrimination2,bing_state_discrimination},
quantum metrology~\cite{metrology}, quantum state and
process tomography~\cite{tomo1,tomo2}, coherent
controls~\cite{konrad}, local filtering
operations~\cite{filtering}, and more recently in quantum
key distribution protocols~\cite{jaskaran_qkd, arvind3}. They are
also significant in
exploring foundational aspects including
Bell non-locality~\cite{nonlocality1,nonlocality2} and
quantum contextuality~\cite{context1} where it is not clear
how these scenarios hold if POVMs are considered instead of
projective measurements.
Especially, in the field of quantum contextuality, where
experimental demonstrations have only been performed for
projective measurements, while experimental signatures of
the same for POVMs is still an ongoing
research~\cite{context_test_robust}. An efficient technique
to perform arbitrary POVMs would greatly benefit these
fields. Photonic systems are
becoming an important platform for quantum information,
communication and computation. Therefore, it is of
tantamount importance to have efficient schemes to realise
POVMs on these systems.

In general, POVMs can be implemented on a quantum system by
coupling the system with an ancillary system and
performing projective measurements on the combined system. POVMs
have  been experimentally realised on optical systems for
quantum state discrimination~\cite{povm_exptt1,povm_exptt2},
quantum state estimation~\cite{povm_exptt4} and entanglement
distillation~\cite{povm_exptt3}. However, these 
schemes require use of large number of beamsplitters (BS), waveplates (WP) and 
other optical elements. It is also not known whether they are optimal or not.
Several theoretical
protocols have also been put forward for their optical
implementation using polarization degrees of freedom of single and 
two-photon~\cite{povm_single,povm_bipartite,bing_operations}, quantum
walks~\cite{povm_quantumwalk} and also on circuit based quantum computers~\cite{povm_gates}. 
There also exist
theoretical protocols to simulate arbitrary POVMs on a
$d$-dimensional quantum systems using classical
randomness and post-selection~\cite{simulating_povm1,simulating_povm2}. 
However, an
efficient experimental implementation of the same has not
yet been done. Further, these theoretical protocols only
succeed probabilistically, with lower success probability
for higher dimensions. Techniques to
implement arbitrary POVMs deterministically for any dimension of
Hilbert space on optical systems is unknown. This limits their applicability in 
quantum information and communication tasks.

In this paper we propose an efficient, scalable and an
all-optical scheme to implement an arbitrary POVM on optical
systems using unitary operations and projective measurements. We use the Naimark dilation theorem and Cosine-Sine (CS)
decomposition to realize the POVMs. Our
scheme requires only $2(n-1)$  number of balanced
BS and other simple linear optical elements to
realize an $n$-outcome POVM. Since all the components
required are linear, this makes our scheme deterministic and
efficient. We explicitly provide simple optical circuits for
the implementation of two-outcome POVMs and symmetric 
informationally complete POVMs (SIC-POVMs) on a
single qubit and provide a method for realizing an arbitrary
mixed state of a photonic quantum system which can be
readily implemented in lab with the current technology. 

This paper is organised as follows: In
Sec.~\ref{Sec:Background} we review the concept of POVMs and the CS
decomposition~\cite{sandeep_csd}. We also discuss in detail the optical setup and the allowed operations on the internal and external degrees of freedom of light.  In Sec.~\ref{Sec:Results} we present our scheme to implement  $n$-outcome POVMs  and the applications of the scheme is presented in Sec.~\ref{sec:applications}. We conclude in Sec.~\ref{Sec:Conclusion}.

\section{Background}
\label{Sec:Background}
In this section, we present the relevant background required for understanding the results in the article. We start with discussing the optical systems considered in this article. Here we discuss the external and internal degrees of freedom of light and the allowed unitary operations on them. We also review the concepts of POVMs and the CS decomposition. 


\subsection{Optical systems}

In this paper we primarily focus on implementing 
POVMs on an optical system using external 
and internal modes of light. Throughout the paper, the  spatial modes of light are regarded as an
external degree of freedom (DoF), while polarization and 
orbital angular momentum (OAM) modes as internal. We present a scheme to implement an $n$-outcome POVM on the internal DoF using spatial modes as ancilla systems. Although, the current scheme works for any internal DoFs with an arbitrary Hilbert space dimensions, we assume that one is capable of 
experimentally implementing unitary operations on any internal DoF of choice.
This is apparent in the OAM DoF where it is still a challenge to experimentally 
implement arbitrary 
local unitaries on them.

Any arbitrary unitary operation $U(N)$ on $N$-spatial modes at can be applied using the prescription 
given by Reck et al.~\cite{ReckZeilinger1994}. In this prescription, $U(N)$ is decomposed using $O(N^2)$ balanced beam-splitters (BBSs) and phase plates. 
The action of a BBS on two spatial modes is represented by the matrix

\begin{equation}
\mathcal{B} = \frac{1}{\sqrt{2}}
\begin{pmatrix}
1 & i\\
i & 1
\end{pmatrix},
\label{eq:bs}
\end{equation}
and the action of a phase plate can be written as
\begin{align}
  \mathcal{P}(\theta) = \begin{pmatrix}
    1 & 0\\
    0 & e^{i\theta}
  \end{pmatrix},
\end{align}
for any arbitrary real parameter $\theta$. One can easily see that any $U(2)$ operator 
$W$ (up to an overall phase) can be realized using three phase plates and two BBSs as
\begin{align}
  W = \mathcal{P}(\theta_1) \mathcal{B} \mathcal{P}(\theta_2) \mathcal{B} \mathcal{P}(\theta_3),
\end{align}
which represents a Mach-Zehnder interferometer. 

The polarization DoF forms a two-dimensional vector space spanned by two orthogonal polarization states. 
All the normal preserving operation on the polarization states span the $SU(2)$ group. Some examples of such operations are half-wave plates (HWP) and quarter-wave plates (QWP), 
whose matrix form in the horizontal and vertical polarization states basis, reads~\cite{jones}
\begin{align}
  H(\theta) &= e^{-\frac{i\pi}{2}}\begin{pmatrix}
    \cos(2\theta) & \sin(2\theta)\\
    \sin(2\theta) & -\cos(2\theta)
  \end{pmatrix},\\
  Q(\theta) &= e^{-\frac{i\pi}{4}}\begin{pmatrix}
    \cos^2\theta +i\sin^2\theta & (1-i)\sin\theta\cos\theta\\
    (1-i)\sin\theta\cos\theta & i\cos^2\theta +\sin^2\theta
  \end{pmatrix}.
\end{align}
Here, $\theta$ is the angle between the fast axis of the WP and the horizontal axis. An arbitrary $SU(2)$ operation on the polarization states can be realized using just one HWP and two QWP~\cite{simon_gadget1,simon_gadget2, arvind1}.

Another interesting DoF of light is the OAM. Laguerre Gauss (LG) modes are the eigenmodes of the paraxial wave-equation, and are represented by two indices $-\infty < \ell < \infty$ and $0 \le p <\infty$. For $p=0$, $\ell$ characterize the orbital angular momentum in the light beam (or single photon)~\cite{Allen_1992}. Such states are known as OAM states of light and have infinitely many orthogonal states. Such states are highly sought after in quantum information and communication tasks due to their large Hilbert space. However, while in theory it is possible to implement 
any arbitrary transformation on OAMs, the methods to realize such transformations are not known.

\subsection{Quantum measurements}
\label{Sec:POVM}
The most general measurements in quantum theory are 
represented by a mathematical model in which the system of interest is 
coupled to a probe via some channel. Later on, a projective measurement is performed on the probe only. Due to the
coupling between the probe and system, it is possible to infer information about the 
state of the latter by a projective measurement on the former. Such models are called quantum instruments~\cite{holevo,heinosaari,wiseman,busch}.

In this paper, we look at a subclass of measurement models having discrete number 
of outcomes in which 
the coupling channel between the probe and system is taken to be a unitary transformation. 
Such measurements are known as POVMs with immense scope of application in 
quantum information, as has been discussed earlier. 

POVMs consists of a set of  $N$
positive operators $\mathcal{M} = \lbrace E_0,
E_1,...,E_{N-1}\rbrace$ known as effects acting on the
$m$-dimensional Hilbert space $\mathcal{H}$, such that
$E_i\geq 0~\forall~i$ and $\sum_{i=0}^{N-1} E_i = \mathds{1}$.  There are three levels of descriptions of a quantum measurement which are the following: 
\begin{itemize}
\item  In the first level of description, only the statistics of the outcomes of the POVM 
are of interest with only the measurement effects $\lbrace E_i\rbrace$ being fully specified. The probability of $i$-th measurement outcome is given by  $p_i \equiv \langle E_i\rangle = {\text Tr} (\rho E_i)$, i.e., the expectation value of the effect $E_i$.

\item The second level of description for quantum measurements involves the state update rules after the measurements. In this level, we specify the measurement operators $K_{ij}$ such that the $i$-th measurement outcome results in the transformation of the system's state
    \begin{align}
\rho' = \frac{\sum_j K_{ij}\rho K^\dagger_{ij}}{\text{Tr}(K_{ij}\rho K^\dagger_{ij})},\label{Eq:l2-POVM}      
    \end{align}
    and the probability for such even is given by $\text{Tr}(K_{ij}\rho K^\dagger_{ij})$. The relation between the effects $E_i$ and the measurement operators $K_{ij}$ reads
    \begin{align}
      E_i = \sum_j K_{ij}^\dagger K_{ij}.
    \end{align}

\item In the third level of description, the detailed interaction between the system and the probe is specified along with the initial state of the probe. If the net effect of this interaction is represented by joint a unitary operator $U_{SP}$ acting on the system and probe, then the general transformation of the system plus probe is written as
    \begin{align}
      \rho'_{SP} = U_{SP}( \rho_S\otimes \sigma_P) U^\dagger_{SP}.
    \end{align}
    Here  the subscripts $S$ and $P$ stand for system and probe, respectively. $\sigma_P = \sum_jq_j\ket{\phi_j}\bra{\phi_j}$ is the initial state of the probe where $\ket{\phi_j}$ and $q_j$ are the eigenvectors and eigenvalues of $\sigma_P$, and $\rho_S$ is the initial state of the system. The application of the unitary operator is followed by a projective measurement on the probe in an orthonormal basis $\{\ket{i}\}$. The measurement operators $K_{ij}$ and the unitary operator $U_{SP}$ is related as
      \begin{align}
        K_{ij} = \sqrt{q_j}\bra{i}U_{SP}\ket{\phi_j}.
      \end{align}
  Hence, the measurement operators $K_{ij}$ contain information about the interaction between the system and the probe as well as  the initial state of the probe.
\end{itemize}

For the special cases, when the probe is in a pure state initially, i.e.,  $\sigma_B = |0\rangle\langle 0|$ the measurement operators become $K_i = \bra{i}U_{SP}\ket{0}$ and the effects read $E_i = K_i^\dagger K_i$. Such measurements are purity preserving measurements as in each measurement outcome the system state is a pure state if initially the system was in pure state. 

Although the state update rule given in Eq.~\eqref{Eq:l2-POVM} is more 
general, for the rest of the paper we deal with measurement operators and their unitary 
equivalents for the choice of pure probe states. This makes some of the calculations easier, while keeping the scheme we present is general enough to also
allow for mixed ancillary states.

Any POVM measurement described according to the second level
with only the measurement operators
$\lbrace K_i \rbrace$ specified, can be realised as a joint unitary
evolution of the system state $\rho_S$
coupled with an ancilla
initialized in the  state $|0\rangle\langle 0|_A$, followed
by a projective measurement on the  latter with  elements
$\lbrace \Pi_i  = \ket{i}\bra{i}\rbrace$, where $\lbrace
|i\rangle \rbrace$ is an orthonormal basis for the finite
dimensional Hilbert space of the ancilla. If the joint
unitary operator is $U$, then the probability to obtain the
$i$-th outcome  on the state $\rho_{AS} = |0\rangle\langle
0|_A\otimes\rho_S$ is then given by,
\begin{equation}
\begin{aligned} P(i) &= \text{Tr}\left(\Pi_i\otimes
\mathds{1} \left[U(|0\rangle\langle 0|_A
\otimes\rho_S)U^\dagger\right]\right)\\
&=\text{Tr}(K_i^\dagger K_i \rho_S), \end{aligned}
\label{eq:naimark} \end{equation}
where $K_i = \langle
i|U|0\rangle$.  Any unitary matrix $U$ of the form
\begin{equation} U=\begin{pmatrix} K_0 &A_{1,1}&\ldots&
A_{1,N-1}\\ K_1 &A_{1,1}&\ldots& A_{2,N-1}\\
\vdots&\vdots&\ddots&\vdots\\ K_{N-1} &A_{N,1}&\ldots&
A_{N,,N-1} \end{pmatrix}, \label{eq:unitary_matrix}
\end{equation}
will result in the measurement operators
$K_i$.  Since we are working in the second level of description of POVMs,
the matrix $U$ is not unique and the $m\times m$ matrices
$A_{i,j}$ can be chosen arbitraily as long as $U$ is
unitary.

Projective measurements are a special case of POVMs where
$E_i$'s are one-dimensional projectors with an additional
constraint Tr$(E_i E_j) = \delta_{ij}$, which states that
the outcomes form an orthonormal basis. Unlike a
non-degenerate projective measurement, the  number of
outcomes in a POVM need not be equal to the dimension of
Hilbert space of the system and can also be continuous. For 
example, a measurement of the direction of spin $1/2$ particles is a POVM with 
a continous spectrum of outcomes. Moreoever, such POVMs can be transformed into  a random 
choice of measurements with a finite number of outcomes~\cite{continous}. However, dealing 
with measurements having a continouos set of outcomes including homodyne and 
heterodyne measurements is beyond the scope of this paper, while 
we primarily focus on discrete outcome measurements only. We implement these
POVMs using unitary transformations and projective measurements.

An important class of POVM is called SIC-POVM. The measurement
effects for SIC-POVMs are proportional to one-dimensional
projectors. For a single qubit the SIC-POVM consists of the
operators $\mathcal{M} = \lbrace\frac{1}{2}\Pi_0,
\frac{1}{2}\Pi_1, \frac{1}{2}\Pi_2,\frac{1}{2}\Pi_3\rbrace$,
where $\Pi_i = |\psi_i\rangle\langle \psi_i|$, such that

\begin{equation}
\text{Tr}(\Pi_i \Pi_j) = \frac{2\delta_{ij}+1}{3}, \quad
\forall i,j \in \lbrace 0,1,2,3\rbrace.
\label{eq:sic_povm}
\end{equation}

The
corresponding measurement operators for this SIC-POVM can be
chosen as $K_i = \frac{1}{\sqrt{2}}\Pi_i$. SIC-POVM are
known to be important for many quantum information
processing tasks~\cite{caves_definneti, fuchs_application}
and quantum measurements~\cite{tomo1, appleby}. For example,
one can perform a full state tomography for a single qubit
state $\rho$ by estimating the probabilities 
$p_i = {\rm Tr}(\rho K_i^\dagger K_i)$ of various
outcomes of the SIC-POVM  given above.
From these outcome probabilities and
information about the SIC-POVM effects $\mathcal{M}$ one can
reconstruct the density operator $\rho$. 
\subsection{CS decomposition}
CS decomposition is a powerful method to decompose an
arbitrary unitary operator $U$ into smaller unitaries and
cosine-sine (CS) matrices~\cite{sandeep_csd}. The most
notable application of this decomposition is in the optical
systems where the smaller matrices correspond to the
operations on the internal degree of freedom of light such
as polarization and OAM and the CS
matrices correspond to generalized Mach-Zehnder interferometers.

CS decomposition states that an arbitrary  $\left(m+n\right)\times \left(m+n\right)$
unitary matrix $U_{m+n}$ ($n\geq m$) can be decomposed into
$n\times n$ and $m\times m$ unitaries and a cosine-sine (CS)
matrix as~\cite{sandeep_csd}

\begin{equation}
U_{m+n} = \begin{pmatrix}
L_{m} & 0\\
0 & L'_n
\end{pmatrix}(\mathcal{S}_{2m}\oplus\mathds{1}_{n-m})\begin{pmatrix}
R^\dagger_{m} & 0\\
0 & R'^{\dagger}_n
\end{pmatrix},
\label{eq:decomp}
\end{equation}
where $\mathcal{S}_{2m}$ is the CS matrix given as,
\begin{equation}\label{Eq:CS}
\mathcal{S}_{2m} = \begin{pmatrix}
C_m & -S_m\\
S_m & C_m
\end{pmatrix}.
\end{equation}
Here $C_m=diag\lbrace \cos\theta_1,
\cos\theta_2,...,\cos\theta_m\rbrace$ and 
$S_m = diag\lbrace\sin\theta_1,
\sin\theta_2,...,\sin\theta_m\rbrace$. The matrix
$\mathcal{S}_{2m}$ can be further simplified as
\begin{equation}
\mathcal{S}_{2m} = \left({B}_2\otimes\mathds{1}_m\right)
\left(\Theta_m\oplus\Theta^\dagger_m\right)
\left({B}^\dagger_2\otimes\mathds{1}_m\right),
\label{eq:bs_internal_decomp}
\end{equation}
where
\begin{align}
  {B}_2 &= \frac{1}{\sqrt{2}}\begin{pmatrix}
1 & i\\
i & 1
\end{pmatrix},\\
  \Theta_m &= {\rm diag}(
\text{e}^{i \theta_1}, 
\text{e}^{i \theta_2}, \cdots, \text{e}^{i\theta_m}).
\label{eq:phase}
\end{align}


Since $L_{m}, L'_{n}, R^\dagger_m, R'^\dagger_n$ are
unitary, the CS decomposition can be applied iteratively to
break the unitaries into smaller dimensional unitaries, as
is explained in detail in Ref.~\cite{sandeep_csd}. In
optical systems, $B_2 \equiv \mathcal{B}$ represents the BBS and the the $\Theta_m$ matrix correspond to the
phase operation on the internal degree of freedom.
For $m=2$, it can be seen that the phase operation corresponds to a
WP.
Hence,
the CS matrix represents a generalized Mach-Zehnder
interferometer. The unitary operations $L_{m}, L'_{n},
R^\dagger_m, R'^\dagger_n$ are the operators acting on the
internal degrees of freedom of light. These unitaries are applied in a
particular path of beam. Hence, they take on the form of control unitaries and are of the form, 
\begin{equation}
L = |0\rangle\langle 0|\otimes L_m + |1\rangle\langle 1| \otimes L'_n,
\end{equation}
with a similar control operator for the unitaries $R^\dagger_m$ and $R'^\dagger_n$. As can be seen the
spatial mode controls which unitary operation acts on the internal degrees of freedom. Such control 
operations can 
be easily handled on optical systems, where the corresponding operators can simply act on the photons 
in different modes locally as shown in Fig.~\ref{fig:fig1}.

\section{Results} 
\label{Sec:Results} 
In this section, we
describe the scheme to implement an arbitrary POVM on
optical systems based on the second level of description of 
quantum measurements in which the measurement operators are 
specified. Here the internal degrees of freedom (DoF)
of photons span the Hilbert space for the systems and the
spatial modes serves as the ancilla. We use Naimark
dilation theorem and CS decomposition to implement the POVM.
In order to implement an $n$-outcome POVM we require $n-1$
BS setups along with unitary transformations on
the internal DoFs in each of the spatial modes, irrespective
of the dimension of the system. As an example, we present a
scheme single shot state tomography
of a quantum state.

We start with the simplest case of POVM, i.e., two-outcome
POVM in Sec.~\ref{Sec:2-out} and generalize this result to
$n$-outcome POVM in Sec.~\ref{sec:n_outcome}. 

\subsection{Two-outcome POVMs}\label{Sec:2-out}

\begin{figure}
\includegraphics[scale=1]{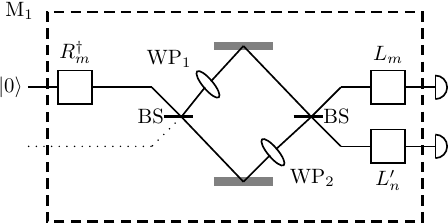}
\caption{An optical setup to implement a two-outcome POVM.
Here the two wires represent the two spatial modes and the
internal degrees of freedom are inherent in each wire. The
operations $R_m^\dagger,L_m$ and $L_m'$   act only on the
internal modes in their respective spatial modes. The
$\mathcal{S}_{2m}$ operator is realized using two BBS and two WPs. Finally, one performs the intensity measurement on
the two spatial modes which yields the probability of the
two measurement outcomes. This optical setup can implement a discrete valued
two-outcome POVMs only.}
\label{fig:fig1}
\end{figure}

Consider a two outcome POVM specified according to the second level of 
description of measurement with measurement operators
$\lbrace K_0, K_1\rbrace$ acting on an $m$-dimensional
Hilbert space. Following the discussion in
sec.~\ref{Sec:POVM} we can write the unitary acting on the
system and a two-dimensional ancilla as 

\begin{equation}
U=\begin{pmatrix}
K_0 & A\\
K_1 & B
\end{pmatrix},
\label{eq:unitary_two_outcome}
\end{equation}
where $A$ and $B$ are appropriately chosen complex matrices
of dimension $m\times m$ such that the matrix $U$ is unitary.
The matrix $U$ is acting on the Hilbert space
$\mathcal{H}_a\otimes \mathcal{H}_s$, where $a$ and $s$
stands for ancilla and system, respectively.

Our aim is to design an optical setup to implement the
unitary operator $U$ and the desired projective
measurements. If we consider the spatial modes of a photon
as the ancilla and the internal modes such as polarization,
orbital angular momentum or frequency modes as the system
states, then using CS decomposition we can decompose $U$
into BS operations on the spatial modes and unitary
operations on the internal modes. 

The CS decomposition of the operator $U$ reads
\begin{equation}
U=\begin{pmatrix}
L_m & 0\\
0& L'_m 
\end{pmatrix}
\begin{pmatrix}
C_m & -S_m\\
S_m & C_m
\end{pmatrix}
\begin{pmatrix}
R^\dagger_m & 0\\
0 & R'^\dagger_m
\end{pmatrix}.
\end{equation}
Here $L_m, L'_m, R^{\dagger}_m$ are $m\times m$ unitary
matrices acting on the internal modes of the photon and
$C_m, S_m$ are cosine and sine matrices [see
Eq.~\eqref{Eq:CS}]. From here we can write 

\begin{equation}
\begin{aligned}
K_0 &= L_m C_m R^\dagger_m,\\
K_1 &= L'_m S_m R^\dagger_m.
\end{aligned}.
\label{eq:svd_two_outcome}
\end{equation}
Eq.~\eqref{eq:svd_two_outcome} 
is simply a singular value
decomposition of the measurement operator $K_i$, which can
be solved  efficiently to get $L_m,L_m',R_m^\dagger$
operators and the $C_m,S_m$ matrices. Furthermore, the
choice of $A$ and $B$ matrices reflects in choosing  the
unitary $R'^\dagger_m$. While working in the paradigm of 
second level of description of quantum measurements in 
which the unitary operation is not completely specified, this freedom can be exploited to
simplify the setup by having  $R'^\dagger_m = \mathds{1}$. We further discuss the 
case of
first and third level of description of quantum measurements in Sec.~\ref{sec:n_outcome}.

The matrix $\mathcal{S}_{2m}$, constructed from  $C_m$ and
$S_m$ can be further decomposed into  BS and phase shift
transformations as given in
Eq.~\eqref{eq:bs_internal_decomp}. 

The local unitary matrices $L_m,L_m',R_m^\dagger$
can be implemented on optical systems and for the case of
polarization degrees of freedom, it requires just one half
WP and two quarter WPs mounted
coaxially~\cite{simon_gadget1,simon_gadget2}.

Finally, a projective measurement is implemented on the 
ancillary modes. This can be viewed as detection of a photon in one or the other 
spatial mode corresponding to the two outcomes of the POVM. The resultant frequency of clicks of 
each detector can then be used to simulate the statistics of the POVM. This way it is possible 
to implement a measurement according to the first level of description. However, it is not 
always necessary to implement a photon detection on both of the spatial modes. The
updated state after the application of the POVM
can be extracted corresponding to a particular outcome by post-selection. As an example, the updated state corresponding to a measurement operator $K_i$, given by Eq.~(6),
can be obtained by not placing any detectors on this mode, while having photon detections
on the modes $j\neq i$. Whenever the detectors on the modes $j\neq i$ do not click,
we are assured to get the required updated state in the $i$th
mode. Using this approach it is thus possible to implement a measurement according to the second and third level of description.

Hence, an arbitrary two-outcome POVM can be implemented in
optical systems. A schematic diagram to implement these POVM
is given in Fig.~\ref{fig:fig1}.

It is to be noted that spatial degrees of freedom (as
denoted by wires in Fig.~\ref{fig:fig1}) are taken as
ancilla while internal degrees of freedom correspond to the
system as in Eq.~\eqref{eq:naimark}. The initialization of
the ancilla state to $|0\rangle\langle 0|$ implies that the
system is injected into the setup through the upper spatial
mode. For a special choice of mixed states of ancilla $\sigma_B = \sum^{1}_{j=0}p_j|j\rangle \langle j|$,
where the eigenstates $|j\rangle$, $j\in \lbrace0,1\rbrace$ corresponds to the upper and lower spatial modes,
the system is injected through the upper spatial mode with probability $p_0$ and 
lower spatial mode with probability $p_1 = 1- p_0$. It should be noted that 
a mixed state can also be prepared probabilistically using our technique described 
in Sec.~\ref{subsec:appl2}. This way we can implement a
general quantum instrument.

\subsection{$n$-outcome POVMs}
\label{sec:n_outcome}

\begin{figure}
\centering
\includegraphics[scale=1]{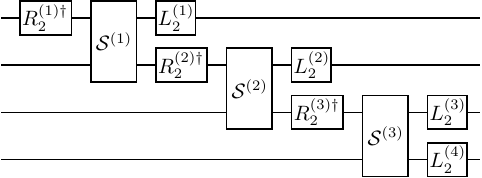}
\caption{A schematic diagram to implement a four-outcome
POVM which can be easily generalized for $n$-outcomes.  Here 
the wires represent spatial modes and the internal DoF are 
inherent in each wire. The
module to implement a two outcome POVM as shown in
Fig.~\ref{fig:fig1} is seen to appear between two different 
modes more than once in this
schematic. This setup can implement a discrete valued four-outcome POVM only.}
\label{fig:4outcome}
\end{figure}

In this section, we generalize the results on the previous
subsection to implement an $n$-outcome POVM on optical
systems.  Consider a POVM with $n$ number of $m$-dimensional
measurement operators $\{K_0,K_1,\cdots K_{n-1}\}$. The
corresponding $nm \times nm$ unitary operator $U$ is given
in Eq.~\eqref{eq:unitary_matrix} with $m\times m$ complex
matrices $A_{i,j}$ chosen such that $U$ is unitary.

Similar to the two-outcome case, we can decompose the matrix $U$  as
\begin{equation}
U=\begin{pmatrix}
L^{(1)}_m & 0\\
0& L'^{(1)}_{m(n-1)} 
\end{pmatrix}
(\mathcal{S}^{(1)}_{2m}\oplus \mathds{1}_{m(n-2)})
\begin{pmatrix}
R^{(1) \dagger}_m & 0\\
0 & R'^{(1)\dagger}_{m(n-1)}
\end{pmatrix}
\label{eq:csd_n_outcome}
\end{equation}
where the unitary operators $L^{(1)}_m$ and
$R^{(1)\dagger}_m$ are of dimension $m$, while
$L'^{(1)}_{m(n-1)}$ and $R'^{(1)\dagger}_{m(n-1)}$ are of
$m(n-1)$ dimension, and $\mathcal{S}_{2m}^{(1)}$ is a
$2\times 2$ block matrix where each block is an $m\times m$
diagonal matrix consisting of cosine and sine as given in
Eq.~\eqref{Eq:CS}. The superscript $(1)$ denotes the first
iteration of CS decomposition.

Since  $A_{i,j}$ are chosen arbitrarily, this gives us
freedom to choose any $m(n-1)$-dimensional unitary for
$R'^{(1)}_{m(n-1)}$. For the sake of simplicity, we choose
$R'^{(1)}_{m(n-1)}=\mathds{1}$. With this choice of
$R'^{(1)}_{m(n-1)}$, we get

\begin{align}
  K_0 &= L_m^{(1)}C^{(1)}_mR_m^{(1)\dagger},\\
  \begin{pmatrix} K_1\\K_2\\\vdots\\K_{n-1}
  \end{pmatrix} &= L_{m(n-1)}'^{(1)}
  \begin{pmatrix}
    S^{(1)}_m\\ \mathcal{O}\\\vdots\\ \mathcal{O}
  \end{pmatrix}R_m^{(1)\dagger}.  
\end{align}
Here $\mathcal{O}$ is an $m\times m$ null matrix. From there
it is clear that only the first $m$ number of columns of
$L_{m(n-1)}'^{(1)}$ will contribute in $K_i$ measurement
operators; therefore, the rest of the columns can be chosen
arbitrarily. In this sense the problem is similar to the one
with $U$ matrix, but now we have $n-1$ measurement
operators.  We can further use CS decomposition for
$L_{m(n-1)}'^{(1)}$ operator to find $m\times m$ unitary
operators $L_m^{(2)}$ and $R_m^{(2)\dagger}$ such that $K_1
=
L_m^{(2)}C^{(2)}_mR_m^{(2)\dagger}S^{(1)}_mR_m^{(1)\dagger}$,
an $m(n-2)$-dimensional unitary operator
$R_{m(n-2)}^{'(2)\dagger} = \mathds{1}$, and an
$m(n-2)$-dimensional $L_{m(n-2)}'^{(2)}$. Recursively CS
decomposing $L_{m(n-j)}'^{(j)}$ unitary operators in the
$j$-th iteration, we can get

\begin{equation}
\begin{aligned}
K_0&= L^{(1)}_mC^{(1)}_mR^{(1)\dagger}_m,\\
K_1 &= L^{(2)}_m C^{(2)}_m R^{(2)\dagger}_m S^{(1)}_m R^{(1) \dagger}_m,\\
K_2 &= L^{(3)}_mC^{(3)}_m R^{(3)\dagger}_mC^{(2)}_m R^{(2)\dagger}_m S^{(1)}_m R^{(1)\dagger}_m\\
\vdots\\
K_{n-2} &= L^{(n-1)}_mC^{(n-1)}_mR^{(n-1) \dagger}_m...S^{(1)}_mR^{(1)\dagger}_m,\\
K_{n-1} &=L^{(n)}_mS^{(n-1)}_mR^{(n-1)\dagger}_m...S^{(1)}_mR^{(1)\dagger}_m.
\end{aligned}
\label{eq:n_outcome_decomposition}
\end{equation}

Since, each of the $\mathcal{S}^{(j)}$ represents a
generalized Mach-Zehnder interferometer which can be
realized using two balanced BS and two diagonal
unitaries on the internal states of photons, an $n$-outcome
POVM can be decomposed into $2n$ general unitary operations
and $2n-2$ diagonal unitaries on the internal states of the
photons and $2n-2$ BBS.

In order to solve Eq.~\eqref{eq:n_outcome_decomposition}, we
can convert these equations into singular value
decompositions as follows: the first equation  is already a
singular value decomposition of $K_0$ which yields
$L_m^{(1)},R_m^{(1)\dagger}$ and the diagonal matrices
$C_m^{(1)}$ and $S_m^{(1)} =
\sqrt{\mathds{1}-\left(C_m^{(1)}\right)^2}$. Using this we
can rewrite the second equation as

\begin{align}
K_1\left(S^{(1)}_m R^{(1) \dagger}_m\right)^{-1} = L^{(2)}_m
C^{(2)}_m R^{(2)\dagger}_m,
\end{align}
which is a singular value decomposition which yields
$L^{(2)}_m, R^{(2)\dagger}_m$ and $ C^{(2)}_m$. Similarly,
one can find all the other matrices and construct the
optical setup. It can be seen that for decomposing
any $K_i$, it is required to evaluate a simple singular value decomposition
problem.

A schematic diagram to implement a $4$ outcome POVM for
$m=2$ is given in Fig.~\ref{fig:4outcome} which can be
extended in a similar fashion to higher number of outcomes
as desired. As can be seen the module to implement a two
outcome POVM appears more than once in the schematic.
A numerical code for the same can be found at~\cite{jorawar}

The most crucial step in solving
Eq.~\eqref{eq:n_outcome_decomposition} is taking the inverse
of matrix $\left(S^{(1)}_m R^{(1) \dagger}_m\right)$. For
the case of $n=2$, this problem does not arise. However, for
$n>2$, the situation is a little non-trivial. For the case
of two-level systems, i.e., $m=2$, the  operator $S^{(1)}_2$
is non-invertible only if $E_0 = K^\dagger_0K_0$ is a rank
one projection. For $m=2$ and $n>2$, not all the $E_i$'s can
be rank one projectors, as can be seen from the condition
$\sum_i E_i = \sum_iK_i^\dagger K_i = \mathds{1}$.
Therefore, we can always find at least $n-2$ number of
measurement operators for which the $S_2$ is invertible.
Therefore, for these systems
Eq.~\eqref{eq:n_outcome_decomposition} can be solved
exactly.

For $m> 2$, solving Eq.~\eqref{eq:n_outcome_decomposition}
may not always be possible. For example, for $m=3$ the
projective measurement contains three measurement operators
each of which are one-dimensional projectors. Therefore, the
first equation in Eq.~\eqref{eq:n_outcome_decomposition}
will yield a rank-1 $C_m^{(1)}$ matrix and rank-2
$S_m^{(1)}$ matrix, which is non-invertible. Therefore, even
for projective measurements it is difficult to solve this
equation using the method prescribed above. 

\begin{figure*}
	\centering
  \includegraphics[scale=1]{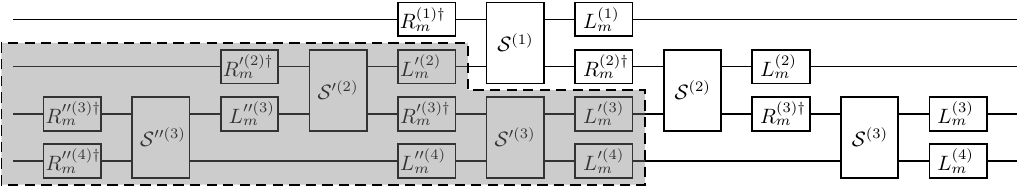}
\caption{A schematic diagram to experimentally realise an
arbitrary $4$ outcome POVM using Gram-Schmidt
orthogonalization method for the non-trivial cases in $m>2$
scenario. The wires represent spatial modes of light, while the 
internal DoF are inherent in each wire. All the optical elements appearing in the grayed out
box can be set to identity without compromising either the
measurement operators or the CS decomposition when the 
either the first or second level of description of a quantum measurement is specified. The resultant
decomposition then resembles the one given in
Fig.~\ref{fig:4outcome}. However,
that is no longer the case when dealing with the third level, when the elements in the grayed box 
can be set to identity. This optical setup can implement discrete valued POVMs only.}
  \label{Fig:CSD}
  \end{figure*}

A numerical method can be used for $m>2$ case. In this
method, we construct the unitary operator $U$ using the
given measurement operators $\{K_i\}$ and choose matrices
$A_{ij}$ randomly such that the matrix $U$ is unitary. This
can easily be ensured by using Gram-Schmidt
orthogonalization. Using CS decomposition numerically on the
operator $U$ we can obtain all the optical components
required to experimentally realize it. Schematically, the
decomposition looks as shown in Fig.~\ref{Fig:CSD}. From
Eq.~\eqref{eq:n_outcome_decomposition} it is known that only
the right most set of $L^{(i)}_m$ operators,
$\mathcal{S}^{(i)}_m$ operators and $R^{(i)\dagger}_m$
operators contribute to the measurement operators $K_i$. All
the rest of the operators in the decomposition can be chosen
arbitrarily, without affecting $K_i$. Therefore, for
convenience we can choose the rest of the operators to be
identity. 

The protocol so presented can accomodate the three levels of 
description of a quantum measurement as described in Sec.~\ref{Sec:POVM}. The 
aforementioned analysis for two and $n$ outcome POVMs is presented keeping in 
mind the second level of description, in which the 
measurement operators are specified. The scheme presenteds in Fig.~\ref{fig:fig1} and Fig.~\ref{fig:4outcome} suffice for this case.

For a description of 
the first level, we are given only the
effects $\{E_i = K_i^\dagger K_i\}$; there can be
infinitely many measurement operators resulting in the same
$E_i$. Any measurement operator $\tilde K_i = WK_i$ where
$W^\dagger W = \mathds{1}$ will also yield $E_i = \tilde
K_i^\dagger\tilde K_i = K_i^\dagger K_i$. In this case we
are free to choose the simplest possible $\tilde{K}_i$ for our
purpose. Using this freedom we can remove the $L_m^{(i)}$
unitary operators from the setup. Therefore, we need only
$n$ number of general unitaries, $2n-2$ diagonal unitaries
and $2n-2$ BBS. This greatly simplifies our experimental setup.

For the third level of description, the unitary operator is completely
specified and therefore, we no longer have the freedom to choose $R'^\dagger_m = \mathds{1}$. This
complicates the setup as more unitaries have to be implemented now. A schematic diagram to implement
a $4$ outcome POVM according to the third level of description is given in Fig.~\ref{fig:4outcome}, where 
the grayed out components can no longer be set to identity. As is evident from the discussion, the 
complexity of the experimental setup increases with the level of description of the POVM.

\section{Applications}
\label{sec:applications}

In this section we detail some applications of CS
decomposition to implement POVMs on optical systems and
perform their cost analysis in
terms of required number of optical
elements. 

\subsection{Single-shot quantum state tomography of photonic qubit}

In general, complete state tomography of a two-level system
requires estimating the expectation value of three
non-commuting observables. This can be done using three
measurement settings. However, changing the experimental
setting can cause errors and misalignment, which can result
in inaccurate outcomes. Single-shot quantum state tomography
is a technique  where a single experimental setup can be
used to estimate the state of a quantum system. This
technique often requires measurements involving POVMs. One
can use SIC-POVM  to perform single-shot state tomography.
However, generally it is difficult to perform four-outcome
POVMs on optical systems.

In order  to implement a SIC-POVM on polarization qubits
using our scheme we need six BS, six WPs and seven unitaries acting
on spatial modes. The schematics of this setup is
given in Fig.~\ref{fig:4outcome}. The $2\times 2$ unitary
operators $L_2^{(i)}$ and $R_2^{(i)}$ can be calculated for
a given  SIC-POVM  using
Eq.~\eqref{eq:n_outcome_decomposition} given in
Sec.~\ref{sec:n_outcome}. The low number of BS
and WP required in our scheme makes it one of  most
viable options to implement single-shot quantum state
tomography in a cost effective manner.

\subsection{Preparing arbitrary  mixed states of a quantum
system} 
\label{subsec:appl2}
Another interesting application of optical
implementation of arbitrary POVM is the preparation of
arbitrary mixed state in optical systems. Mixed state are
important to calibrate the experimental setups. Furthermore,
the mixed states have a fundamental uncertainty as they do
not retain the information about the preparation basis. Here
we show how POVMs can be used to prepare an arbitrary given state
$\rho$ for $m$-dimensional quantum systems.

For this purpose we need a setup for two-outcome POVM and a
maximally mixed state $\rho_r = \mathds{1}/m$ as the input
in the optical setup as shown in Fig.~\ref{fig:fig1}. 
The measurement operators we choose
are $K_0 = \sqrt{\rho/\lambda_0}$ and $K_1 =
\sqrt{\mathds{1}-\rho/\lambda_0}$ where $\lambda_0$ is the
largest eigenvalue of the state $\rho$. One can check that
$K_0^\dagger K_0  + K_1^\dagger K_1 = \mathds{1}$. 

Using the
technique described in Sec.~\ref{Sec:2-out} we can decompose the 
POVM into simple unitaries acting on internal and external DoFs. The choice of local unitaries and phase  shifters will depend on the state $\rho$ that is being prepared and can be easily calculated using  our prescription. At this stage no final projective measurement as described in Sec.~\ref{Sec:2-out} 
has beeen made on any of the 
output modes. Consequently, the output states in each of the two modes can be seen to
correspond to
$K_0\rho_r K_0^\dagger =
\rho/(m\lambda_0)$ with probability $p_0 = 1/(m\lambda_0)$ in the upper spatial mode
and $E_1 = K_1\rho_r K_1^\dagger = (\mathds{1} -
\rho/\lambda_0)/m$ with probability $p_1 =
1-1/(m\lambda_0)$ in the lower spatial mode. Hence, the state of the internal degree of freedom in the upper spatial mode is the desired state $\rho$ before performing the measurement.

Interestingly, for the case of $m=2$ the eigenvalues of the
density operator $\rho$ are of the form $\lambda,
1-\lambda$, for $0\le \lambda \le 1$. The spectrum of
$\mathds{1}-\rho$ is the same. Therefore, we can find a
unitary transformation $V$ such that
$V(\mathds{1}-\rho)V^\dagger = \rho$. Hence, using an
appropriate transformation $V$ on the lower spatial mode, we 
can transform the outcome into $\rho$ too.
This gives us $100\%$
success rate for creating an arbitrary single-qubit mixed
state.

\section{Conclusion}
\label{Sec:Conclusion}
In this paper we detailed a protocol to implement any
arbitrary POVMs on internal degrees of freedom of a
light beam. The basis of our protocol lies in CS
decomposition which can be used to decompose any complicated
unitary matrix into simpler ones. Applying Naimark's
dilation theorem in conjunction with CS decomposition it is
possible to find a unitary matrix corresponding to any POVM
which can then be further decomposed.  Using our method any
$n$ outcome POVM acting on an arbitrary dimension Hilbert
space can be experimentally implemented with a $100\%$
success rate with a far lower number of optical elements
than the current techniques.

Furthermore, our technique can quite easily accomodate the 
three levels of description of quantum measurements by modifying the optical 
setup by incorporating more (or less) local unitaries acting on the internal DoF. We show that the 
complexity of the 
experimental setup increases with the level of description.

Since it is quite hard to experimentally implement POVMs
with arbitrary number of outcomes, our scheme makes it
possible to study their effects and applications in
scenarios like local filtering, state
tomography~\cite{tomo1,tomo2}, quantum key
distribution~\cite{jaskaran_qkd} and quantum
non-classicality~\cite{context1, arvind2}, where they are known to play
an important role.

Further, the techniques discussed in this paper are
experimentally feasible with the current technology and can
be readily implemented.

\begin{acknowledgments}
JS acknowledges support from CSIR-UGC NET, India. JS and A 
acknowledge the funding from the project DST/ICPS/QuST/Theme-1/2019/General Project No. Q-68.. S.K.G.
acknowledges the financial support from Inter-disciplinary
Cyber Physical Systems(ICPS) programme of the Department of
Science and Technology, India, (Grant
No.:DST/ICPS/QuST/Theme-1/2019/12).

\end{acknowledgments}

\end{document}